\documentclass[aps,showpacs,preprintnumbers,amsmath,amssymb]{revtex4}
\usepackage{graphicx}% Include figure files
\usepackage{dcolumn}% Align table columns on decimal point
\usepackage{bm}% bold math
\newcommand{\beq}{\begin{equation}}
\newcommand{\eeq}{\end{equation}}
\newcommand{\beqa}{\begin{eqnarray}}
\newcommand{\eeqa}{\end{eqnarray}}
\newcommand{\beqar}{\begin{eqnarray*}}
\newcommand{\eeqar}{\end{eqnarray*}}
\newcommand{\bra}[1]{\mbox{$\left\langle{#1}\right|$}}
\newcommand{\ket}[1]{\mbox{$\left|{#1}\right\rangle$}}

\newcounter{saveeqn}

\begin{document}
\title{Teleportation of an arbitrary mixture of diagonal states of
multiqubits\\ {$\bm via$} classical correlation and classical
communication}
\author{An Min Wang}
\affiliation{Department of Modern Physics, University of Science
and Technology of China, Hefei, 230026, P.R.China}

\begin{abstract}
We propose a protocol in which the faithful and deterministic
teleportation of an arbitrary mixture of diagonal states is
completed $via$ classical correlation and classical communication.
Our scheme can be generalized straightforwardly to the case of
$N$-qubits by using $N$ copies of classical correlated pairs and
classical communication. Moreover, a varying scheme by using the
generalized classical correlated state within a multiqubit space is
also presented. In addition, the arbitrary mixed state whose set of
eigenvectors is known are a direct application of our protocol.
\end{abstract}


\maketitle

\homepage[]{http://qtg.ustc.edu.cn}
%\altaffiliation{It is founded by the National Fundamental Research
%Program of China with No. 2001CB309310, partially supported by the
%National Natural Science Foundation of China under Grant No.
%60173047  and the Natural Science Foundation of Anhui Province}

\section{Introduction}
\label{intro}

Since Bennett, Brassard, Cr\'{e}peau, Jozsa, Peres, and Wootters
(BBCJPW) proposed quantum teleportation in 1993
\cite{teleportation}, many theoretical protocols were suggested and
some experimental implementations were proposed. Recently, the
teleportation of multiqubits has been well studied \cite{Rigolin}.
Quantum teleportation has become the one of the most important and
the most influential achievements in quantum theory, in particular,
in the end of twenty century.

Quantum teleportation transports an unknown quantum state from Alice
(sender) to Bob (receiver) $via$ quantum correlation
(Einstein-Podolsky-Rosen's pair) and classical communication. When
transporting a known quantum state, one can used the remote state
preparation (RSP) which was proposed by Lo \cite{Lo} and Pati
\cite{Pati}. As to the protocol of transporting a partially known
quantum state, sometimes it is called as teleportation and sometimes
it is called the remote state preparation either. Whatever if the
quantum state one wants to transport is known, partially known or
unknown, it is very interesting and important to know what resources
will be costed at least and which resources can be replaced by the
other ones or be traded off among the used resources, as well as how
new resources can be exploited. This is just our main motivation. In
addition, we would like to show the action of classical correlation
in quantum information processing and further help for understanding
the nature of quantum and classical correlations produced in quantum
theory.

In this paper, we propose a protocol in which the faithful and
deterministic teleportation of an arbitrary mixture of diagonal
states is completed $via$ classical correlation and classical
communication. Here, a classical correlated pair of two qubits can
be written as \beq
C^p_{AB}=\frac{1}{2}\left(\ket{00}_{AB}\bra{00}+\ket{11}_{AB}\bra{11}\right)
\eeq Its name is from that it does not violate local hidden variable
(LHV) theory \cite{Werner}. In fact, it is a separable mixed state
of two qubits and then there is no any quantum entanglement. In
transporting and distributing quantum state, it has played a
substituting and active role. For example, Cubitt $et al.$
\cite{Cubitt} used it to distribute entanglement. Toner and Bacon
\cite{Toner} constructed a protocol showing that the teleportation
of a single qubit admits a local hidden variable theory. Ghosh
$et.al$ \cite{Ghosh} provided an alternative simple proof of the
necessary of entanglement in quantum teleportation and tried to show
that it is sufficient to have a classical correlated channel in
order to teleport any commuting qubits.

Our scheme can be generalized straightforwardly to the case of
multiqubits. Teleporting an arbitrary mixture of diagonal states of
$N$-qubits needs to using $N$ copies of classical correlated pairs
and classical communication. Actually, we also can use the
generalized classical correlated state within a multiqubit space to
carry out our teleportation. In addition, we discuss the application
of our protocol to the arbitrary mixed state whose set of
eigenvectors is known.

\section{One qubit}
\label{sec:1}

First, let us describe how to teleport a mixture of diagonal states
of one qubit $via$ a classical correlation and classical
communication \cite{Our1,Ghosh}. As a class of mixed state, such a
state can be written as \beq \rho^d_X(1)=a_0\ket{0}_X\bra{0} + a_1
\ket{1}_X\bra{1} \eeq where for a density matrix, $a_0$ and $a_1$
ought to be real and positive, as well as their summation is 1.

Initially, Alice and bob shared a classical correlated pair. Thus,
the joint system of them is just \beq \rho_{\rm
ini}(1)=\rho^d_X(1)\otimes C^p_{AB} \eeq where the first two qubits
(denoted by the subscripts $XA$) belong to Alice and the third one
(denoted by the subscript $B$) belongs to Bob. Our protocol can be
divided into two steps.

Step one: Alice performs the operation \footnote{In our notion, we
always keep the structure of Hilbert space and the reduced space to
Alice or to Bob is easy to obtained.} \beq \label{Aliceop} O_{\rm
Alice}(1)=(\sigma_0^X\otimes
H^A\otimes\sigma_0^B).(CNOT_0(X,A)\otimes\sigma_0^B) \eeq where
$\sigma_0$ is $2\times 2$ unit matrix, $H$ is a Hadamard
transformation and $CNOT_0(X,A)$ is a controlled {\sc not} of two
neighbor qubits in which $A$ is a control qubit lying at the latter.
The definitions of these operations are respectively \beq
H=\frac{1}{\sqrt{2}}\left(\begin{array}{cc} 1 &
1\\1&-1\end{array}\right),\qquad
CNOT_0(X,A)=\sigma_0^X\otimes\ket{0}_A\bra{0}+\sigma_1^X\otimes\ket{1}_A\bra{1}
\eeq where $\sigma_1$ is Pauli matrix. The transformation skill
(\ref{Aliceop}) is actually from M. A. Nielsen and I. L. Chuang's
idea \cite{Nielsen} in order to change Bell's basis measurement into
the computation basis measurement. Here, our aim is to make that
Alice can execute the measurement in the computation basis.

It is easy to obtain the transformed result as the following: \beqa
\frac{1}{4}\left(\ket{00}_{XA}\bra{00}+\ket{01}_{XA}\bra{01}\right)\otimes
\left(a_0\ket{0}_B\bra{0}+a_1\ket{1}_B\bra{1}\right)\nonumber \\
+\frac{1}{4}\left(\ket{10}_{XA}\bra{10}+\ket{11}_{XA}\bra{11}\right)\otimes
\left(a_1\ket{0}_B\bra{0}+a_0\ket{1}_B\bra{1}\right) \nonumber\\
+\frac{1}{4}\left(\ket{00}_{XA}\bra{01}+\ket{01}_{XA}\bra{00}\right)\otimes
\left(a_0\ket{0}_B\bra{0}-a_1\ket{1}_B\bra{1}\right) \nonumber\\
+\frac{1}{4}\left(\ket{10}_{XA}\bra{11}+\ket{11}_{XA}\bra{10}\right)\otimes
\left(a_1\ket{0}_B\bra{0}-a_0\ket{1}_B\bra{1}\right) \eeqa

Step Two: Alice executes the measurement in the computation basis:
$\{\ket{00}_{XA}\bra{00}$, $\ket{01}_{XA}\bra{01}$,
$\ket{10}_{XA}\bra{10}$, $\ket{11}_{XA}\bra{11}\}$ and she has the
probability 0.25 of reducing to one of them. For the first two
basis, she can send a cbit 0 to Bob and so Bob has
$a_0\ket{0}_B\bra{0}+a_1\ket{1}_B\bra{1}$ as required. While for the
later two basis, she can send a cbit 1 to Bob and then Bob has to
apply a $\sigma_1$ ({\sc not}) transformation to obtain
$a_0\ket{0}_B\bra{0}+a_1\ket{1}_B\bra{1}$ as required.

\section{Tow qubits}
\label{sec:2}

Now, let us consider the case of an arbitrary mixture of diagonal
states of two quits. The state to be teleport reads \beq
\rho^d_{X_1X_2}(2)=\frac{1}{4}\sum_{x_1,x_2=0}^1
a_{x_1x_2}\ket{x_1x_2}_{X_1X_2}\bra{x_1x_2} \eeq Alice and Bob have
to share two classical correlated pairs, that is, the joint system
of them can be written as \beq \rho_{\rm
ini}(2)=\rho^d_{X_1X_2}(2)\otimes C^p_{A_1B_1}\otimes
C^p_{A_2B_2}\eeq where four qubits $X_1X_2$ and $A_1A_2$ belong to
Alice, and two qubits $B_1B_2$ belong to Bob.

Similar to the case of one qubit, we use two steps to teleport
$\rho_{X_1X_2}^d(2)$ from Alice to Bob.

Step one: Alice performs the operation \beqa O_{\rm
Alice}(2)&=&\left(\sigma_0^{X_1}\otimes \sigma_0^{X_2}\otimes
H^{A_1}\otimes \sigma_0^{B_1}\otimes H^{A_2}\otimes
\sigma_0^{B_2}\right)\nonumber\\
& &.\left(CNOT_1(X_1,A_1)\otimes\sigma_0^{B_1}\otimes
\sigma_0^{A_2}\otimes \sigma_0^{B_2}\right)\nonumber\\
& &.\left(\sigma_0^{X_1}\otimes CNOT_2(X_2,A_2)\otimes
\sigma_0^{B_2}\right) \eeqa where the control {\sc not} of parting
$n$ qubits is defined by \beq
CNOT_n(X_i,A_i)=\sigma_0^{X_i}\otimes\sigma_0^{\otimes n}\otimes
\ket{0}_{A_i}\bra{0}+ \sigma_1^{X_i}\otimes\sigma_0^{\otimes
n}\otimes \ket{1}_{A_i}\bra{1}, \quad (i=1,2;\;n\geq 1) \eeq It is
not difficult to calculate out the transformed state $\rho_{\rm AO}$
\beqa \rho_{\rm AO}&=&\frac{1}{16}\sum_{x_1,x_2=0}^1
\sum_{\alpha_1,\alpha_1^\prime=1}^1
\sum_{\alpha_2,\alpha_2^\prime=1}^1 \sum_{\beta_1,\beta_2=1}^1
a_{\beta_1\beta_2}\ket{x_1x_2}\bra{x_1x_2}\nonumber \\
& & \otimes \ket{\alpha_1}\bra{\alpha^\prime_1}
\otimes\left[(1-\alpha_1)\sigma_0+\alpha_1\sigma_3\right]
\sigma_{x_1} \ket{\beta_1}\bra{\beta_1}\sigma_{x_1}
\left[(1-\alpha_1^\prime)\sigma_0+\alpha_1^\prime\sigma_3\right]
\nonumber \\
& & \otimes\ket{\alpha_2}\bra{\alpha^\prime_2}
\otimes\left[(1-\alpha_2)\sigma_0+\alpha_2\sigma_3\right]
\sigma_{x_2} \ket{\beta_2}\bra{\beta_2}\sigma_{x_2}
\left[(1-\alpha_2^\prime)\sigma_0+\alpha_2^\prime\sigma_3\right]
\eeqa It has 256 terms, however, only 64 diagonal terms are
important because the non-diagonal terms will not appear after
measurement. Obviously, the diagonal terms with the forms as the
following: \beq
\frac{1}{16}\ket{x_1x_2}\bra{x_1x_2}\otimes\sum_{\beta_1,\beta_2=0}^1
a_{\beta_1
\beta_2}\left[\ket{\alpha_1}\bra{\alpha_1}\otimes\left(\sigma_{x_1}
\ket{\beta_1}_{B_1}\bra{\beta_1}\otimes\sigma_{x_1}\right)\otimes
\ket{\alpha_2}_{A_2}\bra{\alpha_2}\otimes\left(\sigma_{x_2}
\ket{\beta_2}_{B_2}\bra{\beta_2}\sigma_{x_2}\right)\right] \eeq
where $x_1,x_2,\alpha_1,\alpha_2=0,1.$

Step Two: Alice executes the measurement in the 16 computation
basis:
$\{\ket{x_1x_2}_{X_1X_2}\bra{x_1x_2}\otimes\ket{\alpha_1}_{A_1}\bra{\alpha_1}
  \otimes\sigma_0^{B_1}\otimes\ket{\alpha_2}_{A_2}\bra{\alpha_2}\otimes\sigma_0^{B_2}$
$(x_1,x_2,\alpha_1,\alpha_2=0,1)$ and she has the probability $1/16$
of reducing to one of them. It is clear that all of computation
basis can be divided in to four groups respectively corresponding to
$\{x_1=0, x_2=0\}$, $\{x_1=0, x_2=1\}$, $\{x_1=1, x_2=0\}$ and
$\{x_1=1, x_2=1\}$. For each group basis, Alice can send a relevant
cbit to Bob and then Bob has to apply a corresponding transformation
listed in the following table to obtain the teleported state
$\rho_{X_1X_2}^d(2)$.

\begin{center}
\begin{tabular}{l|l}
  \hline\hline
  {\bf Alice's measurement} & {\bf Bob's operation} \\
  \hline
  $\ket{00}_{X_1X_2}\bra{00}\otimes\ket{\alpha_1}_{A_1}\bra{\alpha_1}
  \otimes\sigma_0^{B_1}\otimes\ket{\alpha_2}_{A_2}\bra{\alpha_2}\otimes\sigma_0^{B_2}\quad$
  & $\quad \sigma_0\otimes\sigma_0$ \\
  $\ket{01}_{X_1X_2}\bra{01}\otimes\ket{\alpha_1}_{A_1}\bra{\alpha_1}
  \otimes\sigma_0^{B_1}\otimes\ket{\alpha_2}_{A_2}\bra{\alpha_2}
  \otimes\sigma_0^{B_2}\quad$ & $\quad \sigma_0\otimes\sigma_1$ \\
  $\ket{10}_{X_1X_2}\bra{10}\otimes\ket{\alpha_1}_{A_1}\bra{\alpha_1}
  \otimes\sigma_0^{B_1}\otimes\ket{\alpha_2}_{A_2}\bra{\alpha_2}
  \otimes\sigma_0^{B_2}\quad$ & $\quad \sigma_1\otimes\sigma_0$ \\
  $\ket{11}_{X_1X_2}\bra{11}\otimes\ket{\alpha_1}_{A_1}\bra{\alpha_1}
  \otimes\sigma_0^{B_1}\otimes\ket{\alpha_2}_{A_2}\bra{\alpha_2}
  \otimes\sigma_0^{B_2}\quad$ & $\quad \sigma_1\otimes\sigma_1$ \\
  \hline
  \hline
\end{tabular}
\end{center}

\section{$N$-qubits}

The generalization to the case of $N$-qubits is straightforward but
significant. In order to simplify our notions, we do not write
obviously the symbol of direct product and unit matrix. Suppose the
state to be teleported is \beq \rho_{X_1X_2\cdots
X_N}^d(N)=\frac{1}{2^N}\sum_{x_1,\cdots,x_N=0}^1
a_{x_1,x_2,\cdots,x_N}\ket{x_1,x_2,\cdots,x_N}\bra{x_1,x_2,\cdots,x_N}
\eeq Initially, the joint system of Alice and Bob reads \beq
\rho_{\rm ini}(N)=\rho_{X_1X_2\cdots X_N}^d(N)\prod_{i=1}^N
C^p(A_i,B_i) \eeq where $2N$ qubits $X_1,X_2,\cdots, X_N;
A_1,A_2,\cdots, A_N$ belong to Alice, and $N$ qubits
$B_1,B_2\cdots,B_N$ belong to Bob. Note that $N$ copies of classical
correlated pairs $\prod_{i=1}^N C^p(A_i,B_i)$ are shared by Alice
and Bob.

Step one: Alice performs the operation \beq O_{\rm Alice}(N)=
\prod_{i=1}^N H^{A_i}\prod_{i=1}^N CNOT(X_i,A_i) \eeq That is,
taking each $A_i$ as the control qubit, $X_i$ as the corresponding
target qubit, Alice first performs $N$ {\sc cnot} operations; then
for every $A_i$ qubit, Alice always applies a Hadamard
transformation $H^{A_i}$. The diagonal terms of transformed state
becomes \beq \frac{1}{4^N}\ket{x_1x_2\cdots x_N}\bra{x_1x_2\cdots
x_N}\sum_{\beta_1,\beta_2,\cdots,\beta_N=0}^1
a_{\beta_1\beta_2\cdots\beta_N}\left[\prod_{i=1}^N\left(\ket{\alpha_i}\bra{\alpha_i}\right)
\left(\sigma_{x_i}\ket{\beta_i}\bra{\beta_i}\sigma_{x_i}\right)\right]\eeq

Step two: Alice executes the measurements in the $4^N$ computation
basis $\ket{x_1x_2\cdots x_n}\bra{x_1x_2\cdots
x_n}\otimes\ket{\alpha_1\alpha_2\cdots\alpha_N}
\bra{\alpha_1\alpha_2\cdots\alpha_N}$ $(x_1,x_2,\cdots, x_N=0,1;
\alpha_1,\alpha_2,\cdots,\alpha_N=0,1$) and send a relevant cbit to
Bob. Bob performs the corresponding operation
$\prod_{i=1}^N\sigma_{x_i}$ and obtains the teleported state
$\rho_{X_1X_2\cdots X_N}^d(N)$.

Based on our previous scheme, it is easy to draw out its quantum
circuit:

\begin{figure}[h]%[tbp]
\begin{center}
\includegraphics[scale=0.70]{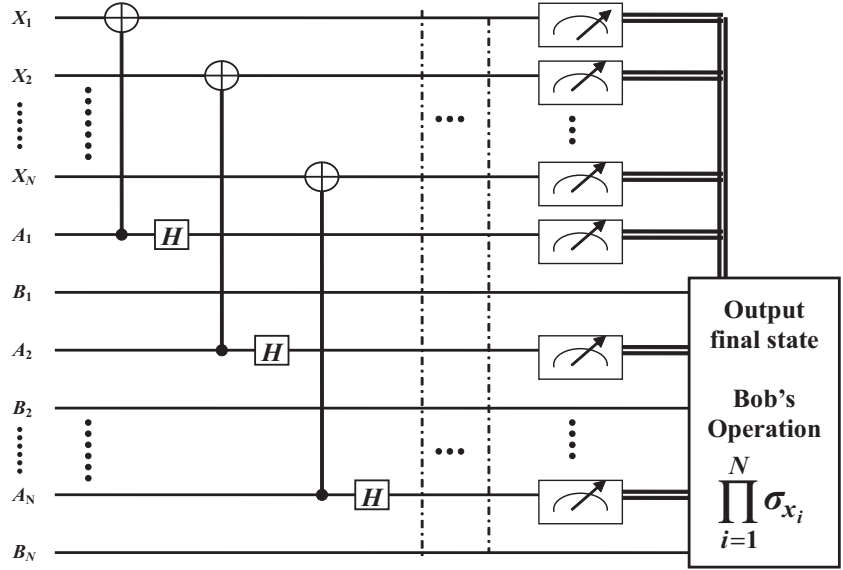} %\qquad
\caption{Quantum circuit of teleportation of an arbitrary mixture of
diagonal states of $N$-qubits}\label{mypic1}
\end{center}
\end{figure}

\section{A varying scheme}

It is worthy of pointing out that there is a varying scheme of our
above protocol by using the generalized classical correlated state
which is constructed by moving all of odd positions of $N$-copies of
classical correlated pairs to the front. In fact, such a
transformation can be completed by a serial of swapping operations.
Introducing a swapping transformation of two neighbor qubits
($2\times 2$ matrix) defined by \beq
S(X,Y)=\left(\begin{array}{cccc}
1 & 0 & 0 &0\\
0& 0 & 1 & 0\\
0& 1 & 0 & 0\\
0& 0 & 0 & 1
\end{array}\right)
\eeq Its action is \beq
S(X,Y)\ket{\alpha_X\beta_Y}=\ket{\beta_Y\alpha_X},\quad
S(X,Y)(Q_X\otimes Q_Y)S(X,Y)=Q_Y\otimes Q_X \eeq Thus, \beq
{\mathcal
S}\left(\prod_{k=1}^N\ket{\alpha_i\beta_i}\right)=\left(\prod_{i=1}^N
\ket{\alpha_i}\right)\left(\prod_{j=1}^N \ket{\beta_j}\right), \quad
{\mathcal S}\left(\prod_{k=1}^N Q_{A_i}Q_{B_i}\right){\mathcal
S}=\left(\prod_{i=1}^N Q_{A_i}\right)\left(\prod_{j=1}^N
Q_{B_j}\right) \eeq where \beq {\mathcal
S}=\prod_{j=1}^{N-1}\left(\prod_{i=1}^{N-j}S(B_i,A_{N-j+1})\right)\eeq
Note that we have used the fact ${\mathcal S}={\mathcal S}^{-1}$.

Therefore, a generalized classical correlated state of $N$-qubit
space can be written as \beq C^s_{A_1A_2\cdots A_N B_1B_2\cdots
B_N}(N)={\mathcal S}\left(\prod_{i=1}^N C_{A_iB_i}^p
\right){\mathcal S} \eeq In the experimental implementation, it is
different from the $N$-copies of classical correlation pairs, and so
it is only not a problem of notion. Just as one considered a
generalized four particle entanglement different from two pairs of
Bell states \cite{Rigolin}. The simplest case is one with two
classical correlated pairs, and the generalized correlated state in
the space of four qubits is just \beqa
C_{X_1X_2A_1A_2}^s(2)&=&\left[\sigma_0^{A_1}\otimes
S(B_1,A_2)\otimes
\sigma_0^{B_2}\right].\left[\frac{1}{2}\left(\ket{00}_{A_1B_1}\bra{00}
+\ket{11}_{A_1B_1}\bra{11}\right)\right.\nonumber\\ & &
\left.\otimes \frac{1}{2}\left(\ket{00}_{A_2B_2}\bra{00}
+\ket{11}_{A_2B_2}\bra{11}\right)\right].\left[\sigma_0^{A_1}\otimes
S(B_1,A_2)\otimes
\sigma_0^{B_2}\right]\nonumber\\
&=&\frac{1}{4}\left(\ket{00}_{A_1A_2}\bra{00}\otimes\ket{00}_{B_1B_2}\bra{00}
+\ket{01}_{A_1A_2}\bra{01}\otimes\ket{01}_{B_1B_2}\bra{01}\right.\nonumber\\
& &\left.+\ket{10}_{A_1A_2}\bra{10}\otimes\ket{10}_{B_1B_2}\bra{10}
+\ket{11}_{A_1A_2}\bra{11}\otimes\ket{11}_{B_1B_2}\bra{00}\right)\eeqa

Suppose that Alice and Bob initially share the generalized state of
classical correlation in the space of $N$-qubits. Thus, the joint
system of Alice and Bob is \beq \rho_{\rm
ini}^s(N)=\rho_{X_1X_2,\cdots,X_N}^d(N)\otimes C^s_{A_1A_2\cdots A_N
B_1B_2\cdots B_N}(N) \eeq Our protocol is then changed as the
following:

Step one: Alice performs the operation \beq O_{\rm Alice}(N)=\left(
\prod_{i=1}^N H^{A_i}\right)\left(\prod_{i=1}^N
CNOT(X_i,A_i)\right){\mathcal S} \eeq That is, only adding a moving
transformation than the previous scheme. In particular, in the case
of two qubits \beqa O_{\rm Alice}(2)&=&(\sigma_0^{X_1}\otimes
H^{A_1}\sigma_0^{X_2}\otimes H^{A_2}\otimes \sigma_0^{B_1}\otimes
\sigma_0^{B_2})\nonumber
\\ & &(CNOT(A_1,X_1)\otimes CNOT(A_2,X_2)\otimes
\sigma_0^{B_1}\otimes \sigma_0^{B_2})\nonumber
\\ & &(\sigma_0^{X_1}\otimes S(X_2,A_1)\otimes\sigma_0^{A_2}\otimes
\sigma_0^{B_1}\otimes \sigma_0^{B_2}) \eeqa

The diagonal terms of transformed state becomes \beq
\frac{1}{4^N}\left(\prod_{i=1}^N\ket{x_i\alpha_i}\bra{x_i\alpha_i}\right)
\sum_{\beta_1,\beta_2,\cdots,\beta_N=0}^1
a_{\beta_1\beta_2\cdots\beta_N}\prod_{j=1}^N\left(
\sigma_{x_j}\ket{\beta_j}\bra{\beta_j}\sigma_{x_j}\right)\eeq For
example, in the case of two qubits it has the form \beq
\frac{1}{16}\ket{x_1\alpha_1x_2\alpha_2}\bra{x_1\alpha_1x_2\alpha_2}
\otimes\left[\left(\sigma_{x_1}^{B_1}\otimes\sigma_{x_2}^{B_2}\right).
\rho_{B_1B_2}^d(2).\left(\sigma_{x_1}^{B_1}\otimes\sigma_{x_2}^{B_2}\right)^\dagger\right]\eeq
where $x_1,x_2,\alpha_1,\alpha_2=0,1.$

Step two: Alice executes the measurements in $4^N$ computation basis
$\prod_{i=1}^N\ket{x_i\alpha_i}\bra{x_i\alpha_i}$ $(x_1,x_2,\cdots,
x_N=0,1$; $\alpha_1,\alpha_2,\cdots,\alpha_N=0,1$) and send a
relevant cbit to Bob. Bob performs a corresponding operation
$\prod_{j=1}^N\sigma_{x_j}$ and obtain the teleported state
$\rho_{X_1X_2\cdots X_N}^d(N)$.

\section{Discussions and conclusions}

Actually, our protocol has more applications. For example, it is
applicable to the teleportation of a general mixed state whose set
of eigenvectors is known (the eigenvalues remain unknown). Alice can
teleport it by first performing a diagonalized unitary
transformation which can be constructed by the set of eigenvectors
of this mixed state, then do our above protocol, and finally, Bob
also has to performs this diagonalized unitary transformation in
order to complete this teleportation.

We have proposed a protocol in which the faithful and deterministic
teleportation of an arbitrary mixture of diagonal states is
completed $via$ classical correlation and classical communication,
however, without quantum entanglement. We show that the
teleportation of an arbitrary mixture of diagonal states of
$N$-qubits needs $N$-copies of classical correlated pairs. It must
be emphasized that the teleported state in our protocol is not fully
unknown but partially known. This is a reason why our protocol does
not need any quantum entanglement. It reminds that there is an
extreme case of trade-off between quantum correlation and classical
correlation under some preconditon. In our point of view, quantum
entanglement is still necessary for a quantum teleportation of a
fully unknown state. A complete quantum teleportation needs both
quantum entanglement and quantum measurement. Our protocol is, at
most, a kind of partially quantum teleportaion since only using
quantum measurement. Just because our protocol depends on quantum
measurement, it is not a classical teleportation one. Of course, the
results presented here offer an intriguing glimpse into the nature
of correlations produced in quantum theory, and show that classical
correlation also is an important and useful resource in quantum
information processing.

\section*{Acknowledgement}

We particularly thank Wan Qing Niu for his earlier work about an
arbitrary mixture of diagonal states of one qubit, and Liang Qiu for
his surveying many related references. We are grateful all of
collaborators of our quantum theory group in the institute for
theoretical physics of our university. This work was founded by the
National Fundamental Research Program of China with No.
2001CB309310, partially supported by the National Natural Science
Foundation of China under Grant No. 60573008.


\begin{references}
%\vskip -0.1in
\bibitem{teleportation} C.H. Bennett, G. Brassard, C. Cr\'epeau, R.
Jozsa, A. Peres, and W. K. Wootters, Phys. Rev. Lett. {\bf 70}, 1895
(1993)
\bibitem{Rigolin}G. Rigolin, Phys. Rev. A {\bf 71}, 032303 (2005)
\bibitem{Lo} H. K. Lo, Phys. Rev. A {\bf 62}, 012313 (2000)
\bibitem{Pati} A. K. Pati, Phys. Rev. A {\bf 63}, 014302 (2001)
\bibitem{Werner}R. F. Werner, Phys. Rev. A {\bf 40}, 4277 (1989)
\bibitem{Cubitt}T. S. Cubitt, F. Verstraete, W. D\"ur, and J. I.
Cirac, Phys. Rev. Lett. {\bf 91}, 037902 (2003)
\bibitem{Toner}B. F. Toner and D. Bacon, Phys. Rev. Lett. {\bf 91},
187904 (2003)
\bibitem{Ghosh}S. Ghosh, G. Kar, A. Roy and U. Sen, Phys. Rev. A
{\bf 65}, 032309 (2002)
\bibitem{Our1} Wan Qing Niu, PhD thesis (2005.4) unpublished.
\bibitem{Nielsen}M. A. Nielsen and I. L. Chuang, ``Quantum
Computation and Quantum Information" (Cambridge University Press,
Combridge, England, 2000)
\end{references}
\end{document}